# A Multi-Venue Solana/DeFi Microstructure Data Corpus: The RED-2400 Family v2

Arati Uday Kamat (ORCID 0009-0000-4781-312X)

---

## Abstract

Empirical research on decentralized and Solana-native market microstructure is constrained less by method than by data: the strongest results in the field rest on a small number of Ethereum-centric datasets, and public, record-level tapes of Solana venues, cross-chain flows, and on-chain oracle behavior remain scarce. We release and document the RED-2400 family v2, a corpus of five public benchmark datasets assembled by a single public-blockchain observer over a concurrent 57-day 2026 window (2026-05-08 to 2026-07-05) under one methodology (free-tier and public feeds only, no paid data, no private accounts, no trading), each with fixed-seed reproducibility scripts, SHA-256 manifests, and a CC-BY-4.0 licence. The five datasets characterize, respectively, Pyth oracle staleness against centralized-exchange prices (164,002 observations), an Aave utilization and Ethereum liquidation panel (18,750 liquidations plus a four-chain utilization series), spot-perpetual basis and funding coupling (244,719 observations), a Solana CEX-versus-DEX spread tape with realized impact (328,186 observations), and a Wormhole cross-chain source-flow tape (360,714 messages). Across the corpus we report three descriptive findings, one honest null, and one field-completeness characterization; none is presented as a causal estimate. The contribution is a coherent, reproducible benchmark family that complements the Ethereum canon and lowers the data barrier for Solana and cross-chain microstructure research.

---

## 1. Introduction

Decentralized finance has produced a rich set of research questions faster than it has produced the datasets needed to answer them. The most credible empirical results in cryptocurrency market microstructure, decentralized-exchange price formation, and DeFi lending rest on a comparatively small number of carefully constructed datasets, and those datasets are overwhelmingly Ethereum-centric. Barbon and Ranaldo (2024), in Management Science, quantify the price efficiency of centralized versus decentralized venues; Lehar and Parlour (2025), in the Journal of Finance, characterize the Uniswap automated market maker; and Capponi, Jia, and Yu (forthcoming, Review of Financial Studies) formalize price discovery on decentralized exchanges. This body of work establishes an Ethereum canon: a shared vocabulary, a set of stylized facts, and a benchmark against which new venues can be measured.

What is thin, by comparison, is the corresponding empirical base for Solana-native venues and for the cross-chain and on-chain-infrastructure layers that increasingly mediate on-chain capital. Solana carries a large and growing share of on-chain trading, yet public record-level tapes of its CEX-versus-DEX price relationships are rare. Price oracles such as Pyth are load-bearing infrastructure for every on-chain derivative and lending market, yet public panels pairing their published prices against a continuously trading benchmark are rarer still. Cross-chain bridges move a large share of capital between ecosystems, but public message-level tapes of individual bridge transfers are scarce and, where they exist, are framed around security rather than flow. The result is an asymmetry: a researcher who wishes to replicate an Ethereum microstructure result on Solana, or to ask a question that spans the oracle, lending, derivative, spot, and bridge layers at once, must typically assemble the data from scratch.

This paper addresses that asymmetry with a data contribution rather than a single empirical thesis. We release and document the RED-2400 family v2, a corpus of five public datasets that share a single provenance, a single collection window, and a single methodology, and that together span five distinct layers of the Solana and DeFi microstructure stack:

1. **RED-ORACLE** — Pyth Network oracle staleness and its co-movement with drift from centralized-exchange prices, for BTC, ETH, and SOL.
2. **RED-LIQ** — an Aave V3 Ethereum liquidation record paired with a four-chain lending-utilization panel.
3. **RED-BASIS** — spot-perpetual basis and funding-rate coupling for major perpetuals.
4. **RED-ARB** — a Solana CEX-versus-DEX spread tape with a realized price-impact figure per observation.
5. **RED-BRIDGE** — a source-side Wormhole cross-chain message tape.

The unifying claim of this paper is deliberately methodological, not empirical. We do not argue that these five phenomena share a single mechanism; they do not. We argue instead that they can and should be measured to a common standard, that doing so lowers the barrier to Solana and cross-chain microstructure research, and that a corpus assembled by one observer over one window under one reproducibility discipline is more useful, and more comparable across layers, than five datasets assembled piecemeal. In this respect the corpus follows an emerging model of individual-researcher benchmark families with companion papers and open reproducibility, exemplified in a different domain by the ForesightFlow prediction-market datasets of Nechepurenko (2026).

Our empirical contribution is correspondingly modest and honest. For each dataset we report descriptive and associational statistics only. Three of the five yield a substantive descriptive finding: oracle staleness co-moves positively with drift from the centralized-exchange price (Pearson correlation about 0.28, with bootstrap confidence intervals excluding zero); Solana DEX prices sit below the centralized benchmark in roughly nine of ten observations, with winning-quote concentration high even across thirty-two venues; and cross-chain bridge flow is strongly Solana-originated. One dataset yields an honest null: short-horizon basis-funding coupling does not predict subsequent funding-regime shifts. One yields primarily a data-quality characterization: the bridge tape's destination, asset, and status fields are largely unpopulated, which bounds what it can support. We are explicit throughout that none of these is a causal estimate; several of the datasets, by construction, do not admit one.

The remainder of the paper is organized as follows. Section 2 situates the corpus in the related literature, grouped by the five layers it spans. Section 3 describes the common provenance, methodology, and reproducibility standard, and summarizes the five datasets. Sections 4a through 4e document each dataset in turn, re-anchored to the shared methodology of Section 3. Section 5 draws the cross-dataset observations together. Section 6 states the corpus's limitations, and Section 7 concludes with directions for follow-up work, including the causal-identification questions this descriptive corpus is designed to enable but does not itself answer.

---

# 2. Related literature

We group the related work by the five layers the corpus spans, and close with the structural precedent for the corpus model itself.

**Centralized-versus-decentralized price formation.** The Ethereum canon on CEX-DEX price relationships is the natural analog against which the Solana tape in this corpus can be read. Barbon and Ranaldo (2024) quantify the market quality of centralized and decentralized venues and show that decentralized prices are less efficient, with gas costs the dominant friction. Lehar and Parlour (2025) characterize the economics of the Uniswap automated market maker and the conditions under which it dominates a limit-order market. Capponi, Jia, and Yu (forthcoming) model price discovery on decentralized exchanges and the role of blockchain fees and informed trading. Oz and colleagues (2025) measure cross-chain arbitrage as the emerging mechanism of price alignment across venues. Relative to this literature, RED-ARB contributes a Solana rather than Ethereum focus, a record-level tape rather than summary statistics, and a realized price-impact field per observation.

**Oracle risk and accuracy.** On-chain price oracles are the infrastructure layer on which every on-chain derivative and lending market depends, and a growing empirical literature measures their behavior. Cong, Prasad, and Rabetti (2025, NBER Working Paper 33639) study financial and informational integration through oracle networks and its association with on-chain economic activity. Gansauer, Ben Aoun, Droll, and Hartenstein (2025) measure Chainlink price-feed accuracy across eight networks against centralized-exchange references using fifty million data points, and document network-dependent deviations under volatility. Nadler, Schuler, and Schar (2024) analyze oracle accuracy and violation recovery, and Ormer (2024) proposes a manipulation-resistant pricing oracle. RED-ORACLE complements this work with a continuous, high-frequency panel of Pyth staleness for three assets, and reports staleness-drift co-movement as an association rather than a causal effect, because publish lag is definitionally the difference between sample time and publish time and admits no instrument.

**DeFi lending and liquidations.** The empirical literature on decentralized lending is anchored by Qin, Zhou, Gamito, Jovanovic, and Gervais (2021), the first large-scale study of DeFi liquidations, and by Cornelli, Gambacorta, Garratt, and Reghezza (2025), who study the motivations of Aave V2 users. Recent work extends this to cross-chain and cross-version settings: Fan and Sun (2025) release a six-chain Aave event infrastructure; Iftikhar, Wei, and Cartlidge (2025) compare automated risk mechanisms across Aave and Compound; Sevim (2026) extends lending-risk models to multi-chain environments; Zbandut and Goldstein (2025) study the migration of risk to a permissionless curator layer; Gadzinski and Liuzzi (2025) find that liquidations do not discourage subsequent borrowing; and Chiu and Danisman (2026) study returns, leverage, and liquidation risk on Aave V3 for the Bank of Canada. RED-LIQ is a task-specific complement to this body of work, and in particular to Fan and Sun (2025): it provides a pre-joined utilization-plus-liquidation panel that reaches into the post-Aave-V3.3 era of 2026.

**Perpetual basis and funding.** The pricing of perpetual futures and the behavior of funding rates is the domain of RED-BASIS. He, Manela, and Ross (2023) derive no-arbitrage prices for perpetuals and document that crypto deviations exceed those in foreign-exchange markets; Kim and Park (2025) study funding-rate design; and Dai, Li, and Yang (2025) examine persistent perpetual-spot discrepancies using two years of Binance data. Schmeling, Schrimpf, and Todorov (2024) document that crypto carry reaches very high levels and predicts liquidations of short futures positions, and Zhivkov (2026) shows that cryptocurrency funding-rate price discovery runs from centralized to decentralized venues. This literature motivates the coupling question RED-BASIS asks, and the honest null it reports: the short-horizon coupling between basis and funding does not predict funding-regime shifts, and a permissive shift definition saturates the base rate.

**Cross-chain bridges.** The closest prior work to RED-BRIDGE is the multi-bridge measurement study of Yan and colleagues (2025, ACM ASIA CCS), and the interoperability-economics study of Cao, Zheng, Cong, Li, and Wang (2026, ACM SIGMETRICS), both of which are multi-bridge and security- or economics-framed. Augusto and colleagues (2024, XChainWatcher) provide an attack-detection dataset. RED-BRIDGE is Wormhole-dedicated and flow-framed, and we are explicit that its analysis-ready fields are the source chain and the USD amount only.

**The corpus model.** Finally, the RED-2400 family follows a structural precedent that is not topical but methodological: the model of an individual researcher releasing a coherent family of public datasets, each with a companion paper, open reproducibility scripts, and a permissive licence. Nechepurenko (2026), in the prediction-markets domain, exemplifies this model with the ForesightFlow pmxt-* dataset family. We adopt the same publication pattern for Solana and DeFi microstructure.

---

# 3. Unified data corpus overview

**Provenance.** All five datasets were assembled by a single public-blockchain observer over a concurrent 57-day window in 2026 (2026-05-08 to 2026-07-05; the dataset-specific start times fall within a roughly six-hour range on the first day). Every value derives from public on-chain data and free-tier or public centralized-exchange endpoints. No paid data feeds, no private or authenticated accounts, and no trading of any kind were used in the construction of any dataset. Transaction hashes and on-chain addresses that appear in the datasets are public identifiers and are not personally identifiable information; a pre-release scan confirmed that no producer, brand, or account label is embedded in any file.

**Common methodology.** The five datasets share a single processing and release discipline. Each is a newline-delimited JSON tape with a frozen v1.0 schema, documented per-field in a companion SCHEMA file. Each ships with a fixed-seed, dependency-light reproducibility script (Node.js or Python) that regenerates the exact headline statistics reported in this paper from the released file, together with a results file recording those statistics. Each ships with a SHA-256 manifest for integrity verification, a CITATION file, and a CC-BY-4.0 licence for the data and an MIT licence for the code. Where a random procedure is used (bootstrap confidence intervals, resampling), the seed is fixed and recorded. All reported statistics in Sections 4 and 5 trace to the deposited results files.

**Framing.** We report descriptive and associational statistics only. Several datasets do not admit a causal design by construction: oracle publish lag is definitionally the difference between sample and publish times; the arbitrage tape lacks signed trades and per-block liquidity depth; and the bridge tape's destination attribution is incomplete. We flag small-sample cases explicitly and do not convert descriptive associations into effect estimates.

**The five datasets.** Table 1 summarizes the corpus. Each dataset is deposited separately on Zenodo with its own concept DOI, cited here as a placeholder pending deposit.

| Dataset | Source | Window (2026) | Size | Primary object | Zenodo record |
|---|---|---|---|---|---|
| RED-ORACLE | Pyth Network + OKX | ~57 days | 164,002 obs | oracle staleness vs CEX drift | 21211110 |
| RED-LIQ | Aave V3 (Ethereum) + 4-chain utilization | 2023-24 + 2026 | 18,750 liq + 2,694 util | liquidations + cross-chain utilization | 21211303 |
| RED-BASIS | OKX perpetuals + spot | ~57 days | 244,719 obs (+2 files) | spot-perp basis and funding coupling | 21211415 |
| RED-ARB | OKX + 32 Solana DEX venues | ~57 days | 328,186 obs | CEX-DEX spread + realized impact | 21211187 |
| RED-BRIDGE | Wormhole | ~57 days | 360,714 msgs | cross-chain source-side flow | 21211498 |

**Table 1.** The RED-2400 family v2. Each Zenodo record identifier resolves to a full DOI of the form `10.5281/zenodo.{record}` (e.g. `10.5281/zenodo.21211110` for RED-ORACLE); all five DOIs are also listed in full in Section 8 (Data availability). All five datasets share provenance, methodology, and release discipline; each is independently deposited and citable. Sizes are the primary-file record counts; RED-LIQ and RED-BRIDGE include a secondary panel/summary file, and RED-BASIS includes a funding cross-section and a term-structure file.

# 4. The five datasets

Each subsection documents one dataset: its structure, its empirical observations, and the boundary of what it supports. All statistics trace to the deposited results file named in Section 3 and to the reproducibility script in the dataset's `analysis/` folder. Figures are the curated subset staged in `figures/`. We flag, at the end of each subsection, the structural feature that recurs across the corpus and is drawn together in Section 5.

## 4a. Oracle staleness panel (RED-ORACLE)

RED-ORACLE is a 164,002-observation panel that pairs, at each sampled instant, the Pyth Network on-chain price for BTC, ETH, and SOL with the contemporaneous OKX price, and records the oracle's publish lag (`lag_sec`, the difference between the sample time and the Pyth publish time) and its signed drift from the centralized price (`drift_bps`). The panel is continuous over the roughly 57-day window and is the corpus's window into the oracle infrastructure layer on which every on-chain derivative and lending market depends.

Publish lag is short in the median but heavy in the tail. The median lag is 2 seconds, but 21.4 percent of observations are at least 5 seconds stale, and a small number of episodes exceed a minute (Figure 4a.1). Over the window we observe 21 episodes with a lag above 60 seconds, the longest reaching 213 seconds (Figure 4a.3). A price that is momentarily stale is, mechanically, a price that can differ from the continuously trading benchmark, and the data bear this out: the magnitude of staleness co-moves positively with the magnitude of drift. Across all three assets the Pearson correlation between lag and absolute drift is approximately 0.28, and the 95 percent bootstrap confidence intervals (fixed seed) exclude zero for each asset. Sorting observations into lag deciles, mean absolute drift rises monotonically across deciles for all three assets (Figure 4a.2).

We are explicit that this is an association, not a causal estimate, and that the dataset does not admit a causal design. Publish lag is definitionally the difference between the sample time and the publish time, so there is no instrument that shifts lag without also shifting the sampling relationship; the co-movement we report is a description of the joint distribution, not an effect of staleness on drift. Within that boundary, the panel establishes a clean, reproducible fact: oracle staleness and price drift are positively and significantly associated at high frequency across the three dominant assets. The one-sided, directional character of that relationship, drift growing with staleness rather than fluctuating symmetrically, is the feature we return to in Section 5.

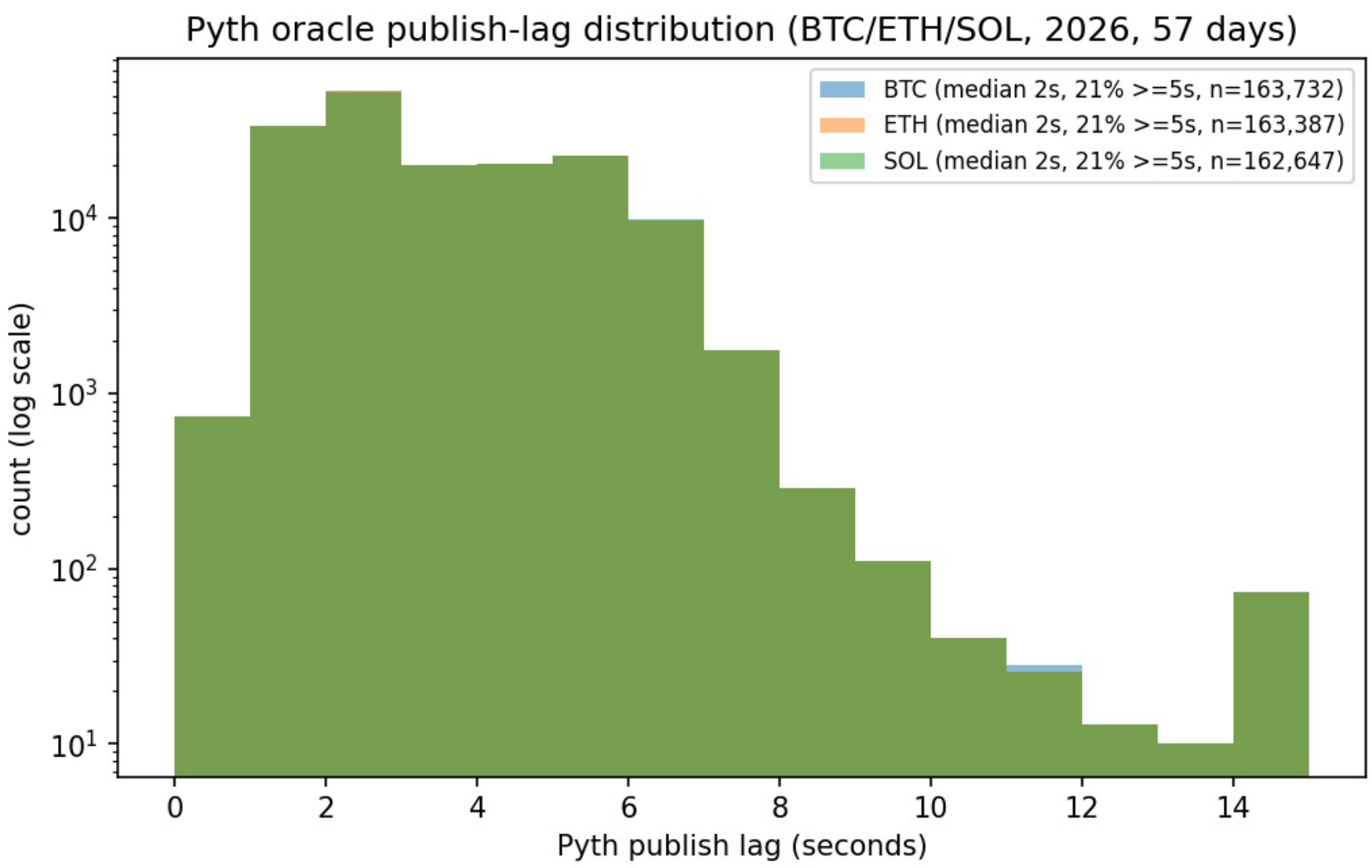

Figure 4a.1. Pyth publish-lag distribution for BTC, ETH, and SOL (log count); median 2 seconds, about 21 percent of prints at least 5 seconds stale.

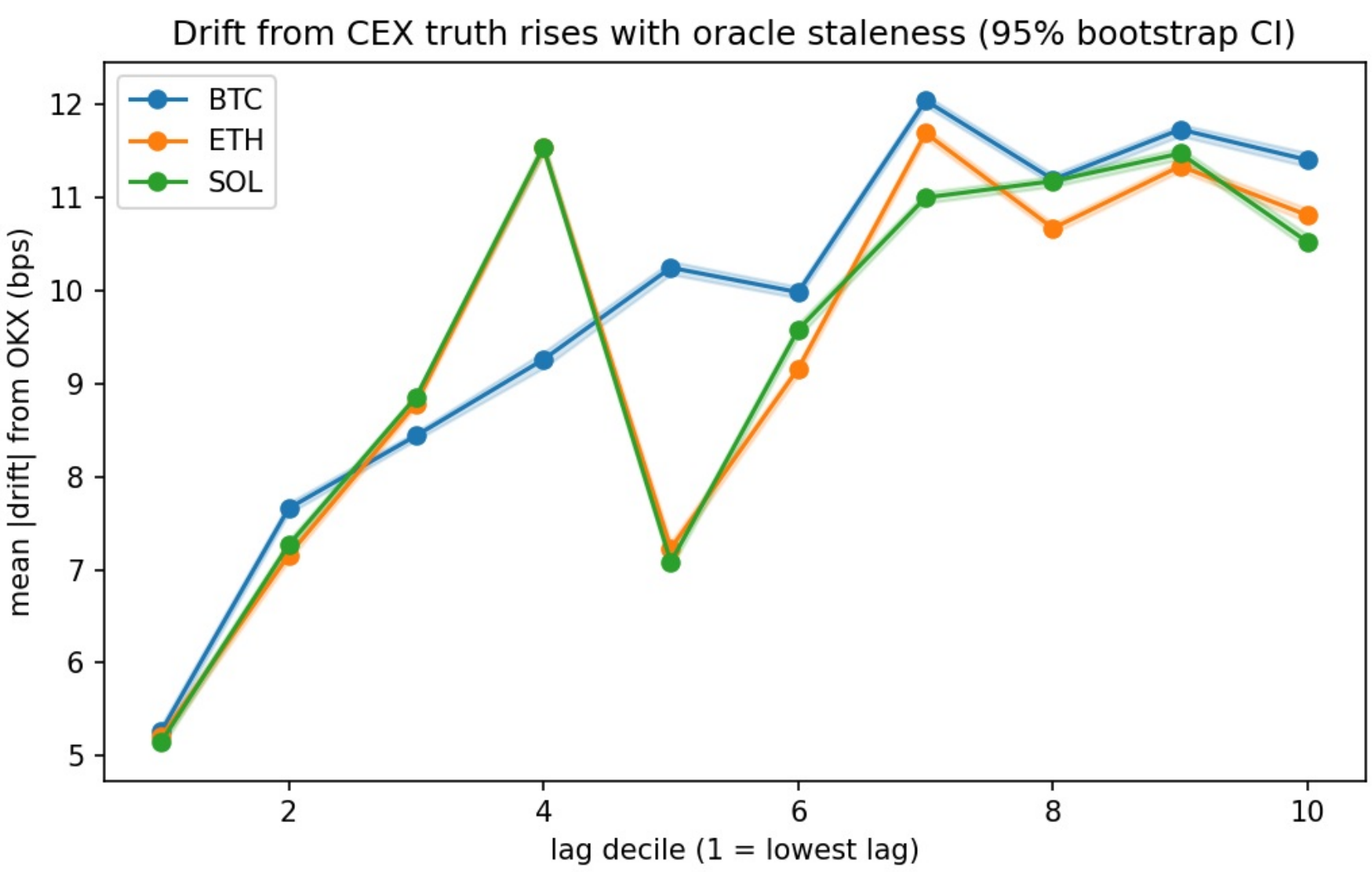

Figure 4a.2. Mean absolute drift from the OKX price by lag decile, with 95 percent bootstrap confidence bands; drift rises monotonically with staleness (correlation about 0.28).

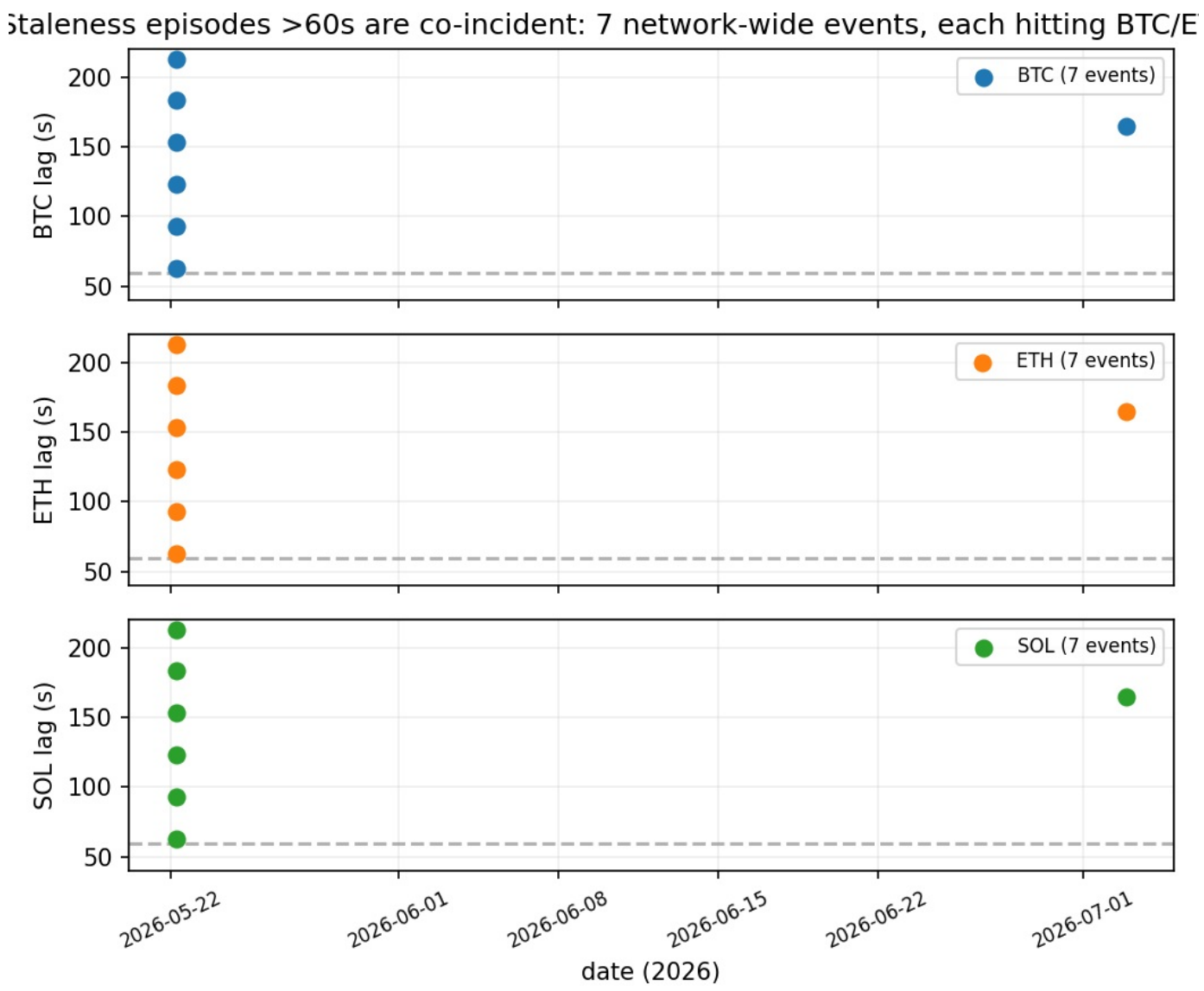

Figure 4a.3. Extreme staleness episodes are co-incident across BTC, ETH, and SOL: seven simultaneous network-wide events, each affecting all three assets (21 asset-episode observations total; maximum 213 seconds). Shown as three stacked panels sharing a common time axis.

## 4b. Solana CEX-versus-DEX spread tape (RED-ARB)

RED-ARB is a 328,186-observation tape that pairs, for six Solana-traded assets, the OKX price with the best available quote across thirty-two Solana decentralized-exchange venues, and records the signed spread (`spread_bps`), the winning venue (`winning_dex`), and a realized price-impact figure per observation (`dex_price_impact_pct`). It is the corpus's window into Solana spot microstructure, and the analog against which the Ethereum CEX-DEX canon can be read.

Three descriptive facts stand out. First, the relationship between the decentralized and centralized price is strongly one-sided: the DEX best quote sits below the OKX price in approximately 91 percent of observations, with a winsorized median spread of about minus 13 basis points (Figure 4b.1). This is a persistent offset, not symmetric dispersion around parity. Second, realized impact at the reference notional is very small, a median of 0.001 percent and a 99th percentile of 0.006 percent, which indicates that the spread, rather than slippage, is the dominant component of the observed gap; we stress that this is an observation tape, not an execution log, and that identifying a genuine arbitrage requires per-block liquidity depth and fee state that are out of scope, so no arbitrage-profit or PIN/VPIN claim is made. Third, and despite the presence of thirty-two venues, the supply of the winning quote is highly concentrated: the winning-venue Herfindahl-Hirschman Index is 0.22, the top five venues supply 80 percent of best quotes, and a single venue, Whirlpool, supplies 42 percent (Figure 4b.2). One asset, JUP, is an outlier, with the DEX above the CEX in 23 percent of its observations against 3 to 8 percent for the others (Figure 4b.3).

The combination of a one-sided price offset and a concentrated winning-venue distribution is the clearest single instance of the structural motif this corpus documents, and we develop it in Section 5.

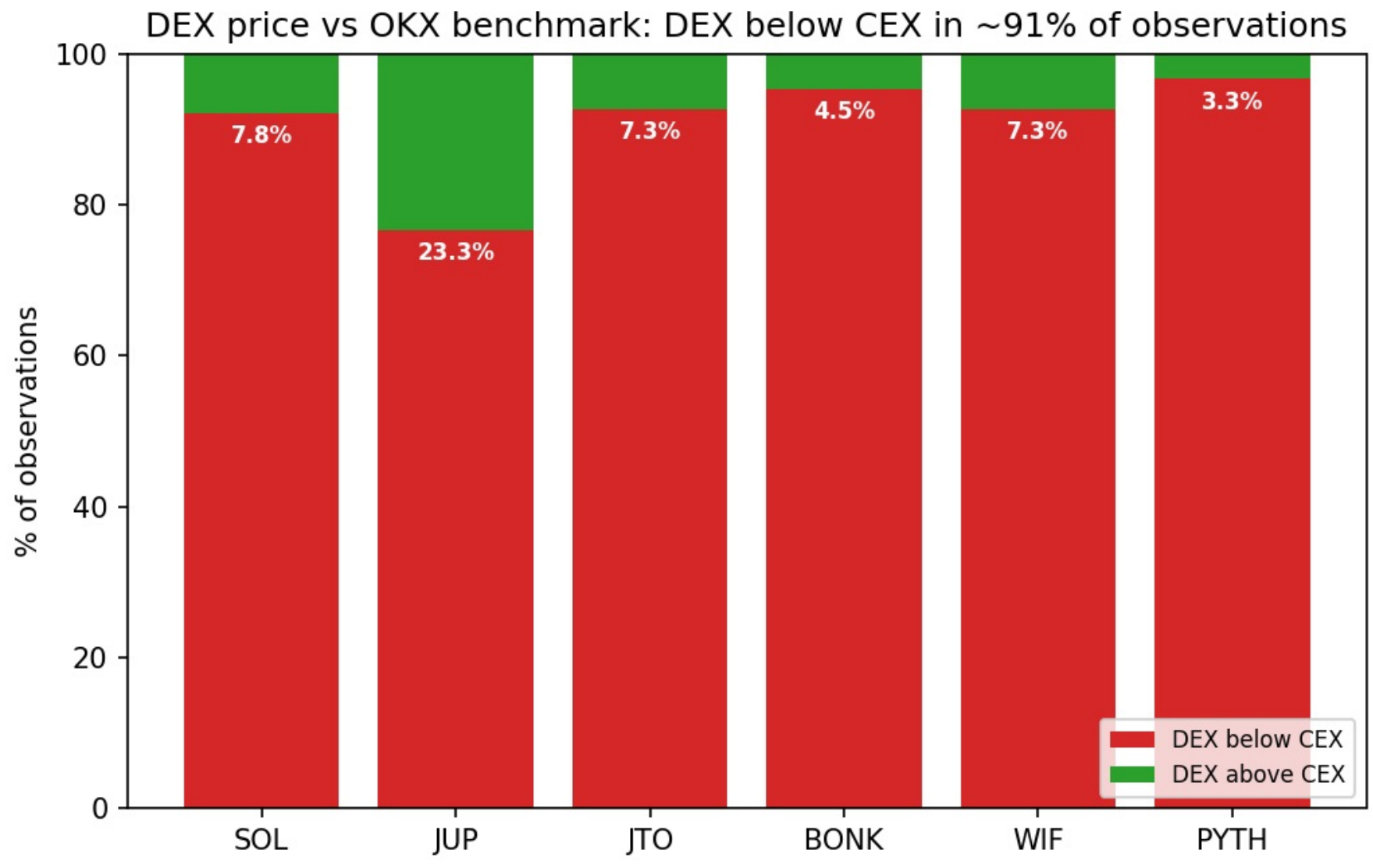

Figure 4b.1. Share of observations with the DEX price above versus below the OKX benchmark, per symbol; the DEX sits below in about 91 percent of observations.

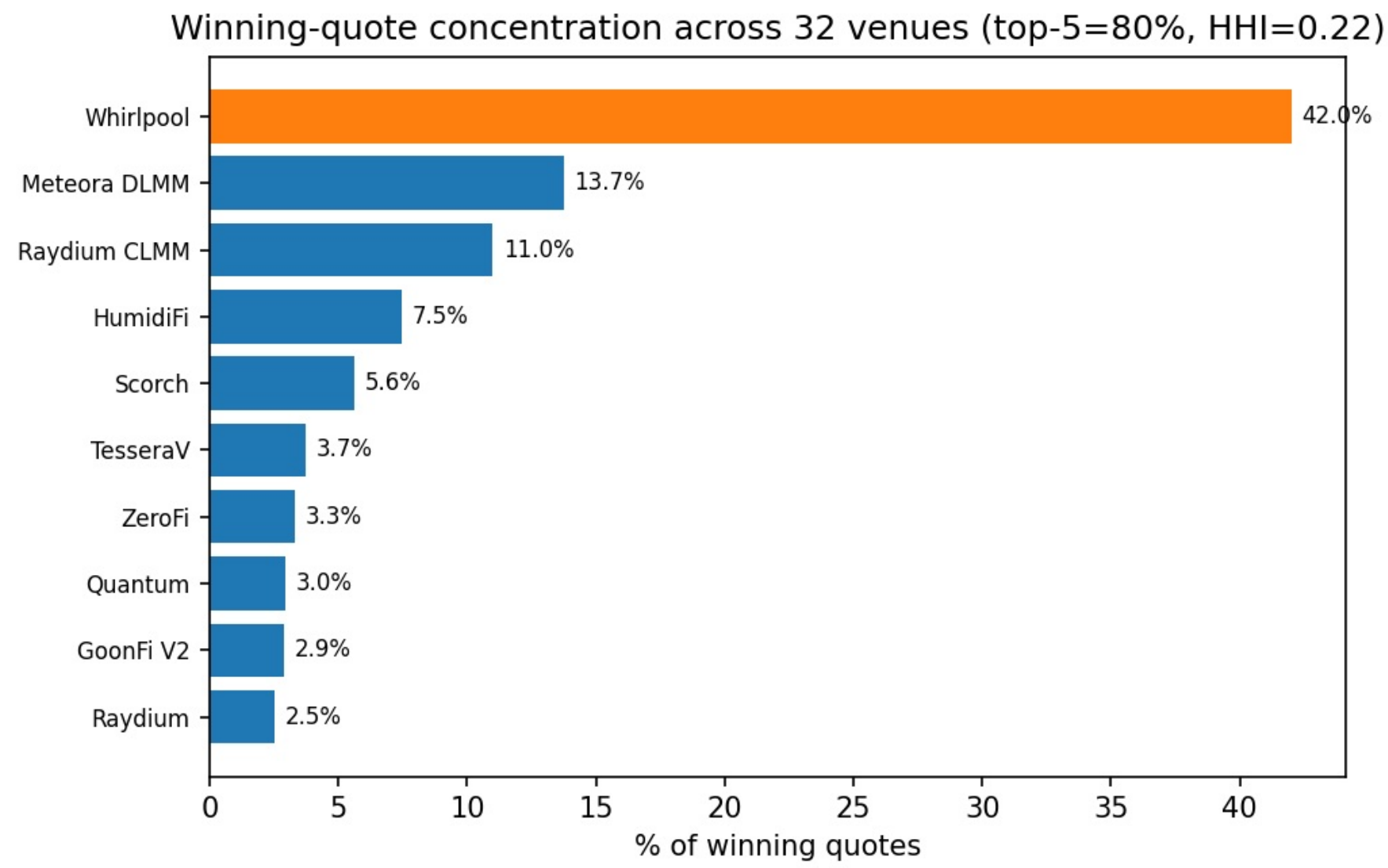

Figure 4b.2. Winning-quote share across the 32 sampled venues; Whirlpool supplies 42 percent, the top-5 supply 80 percent (HHI 0.22).

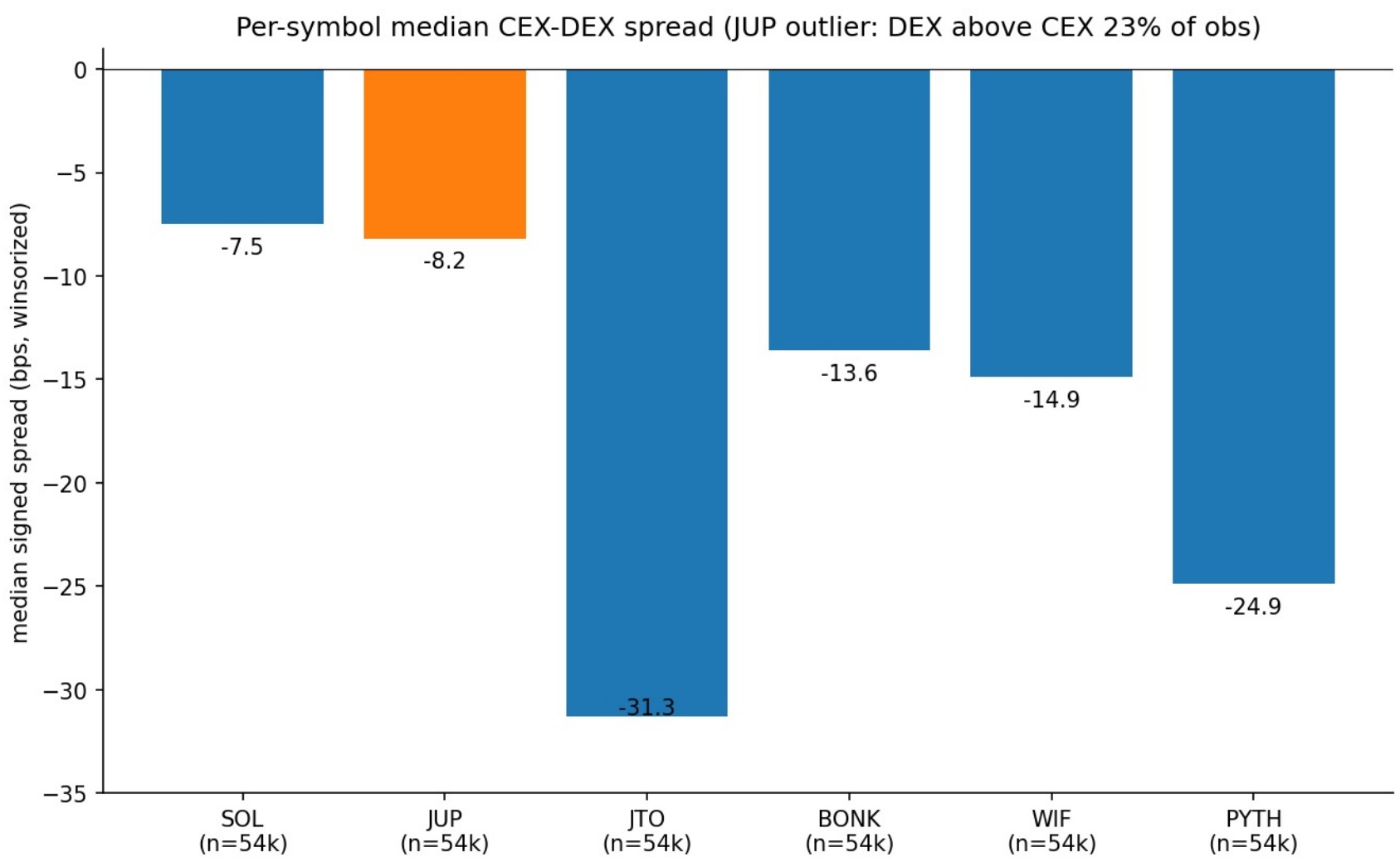

Figure 4b.3. Per-symbol median signed CEX-DEX spread across six Solana tokens. SOL has the smallest absolute spread (−7.5 bps); JUP is the outlier not by spread magnitude (−8.2 bps) but by frequency of DEX-above-CEX prints (23 percent of observations vs. roughly 8 percent for the other five symbols).

## 4c. Aave utilization and liquidation panel (RED-LIQ)

RED-LIQ pairs an Ethereum Aave V3 liquidation record of 18,750 liquidation calls with a four-chain lending-utilization panel of 2,694 observations covering Ethereum, Arbitrum, Base, and Polygon. It is the corpus's window into the DeFi lending layer, and a task-specific complement to the broader Aave event infrastructure of Fan and Sun (2025). The liquidation record spans 2023, 2024, and 2026, with a gap in 2025 (Figure 4c.3); of the 18,750 liquidations, approximately 3,750 fall in 2026 and thus inside the utilization window, forming the analysis-ready paired subset. We document this multi-year structure openly, because it bounds which cross-panel questions the data support.

Two descriptive features are notable. First, liquidation activity is highly concentrated among liquidators: the liquidator Herfindahl-Hirschman Index is 0.071, the Gini coefficient of the liquidator distribution is approximately 0.83, and the top five liquidators account for 46 percent of all liquidations (Figure 4c.1). A small set of specialized actors intermediates a disproportionate share of the liquidation flow. Second, cross-chain utilization is not uniformly synchronized: Arbitrum utilization moves counter to both Ethereum (correlation approximately minus 0.19) and Polygon (correlation approximately minus 0.31), while Ethereum and Polygon themselves co-move positively (correlation approximately plus 0.67). Arbitrum is thus the L2 outlier against a broader Ethereum-Polygon co-cycle (Figure 4c.2). This Arbitrum counter-cyclicality is the panel's headline descriptive finding and is a candidate object for follow-up work.

The liquidator concentration is a second instance, in an entirely separate layer, of the concentration motif seen in the arbitrage tape, and we join the two in Section 5.

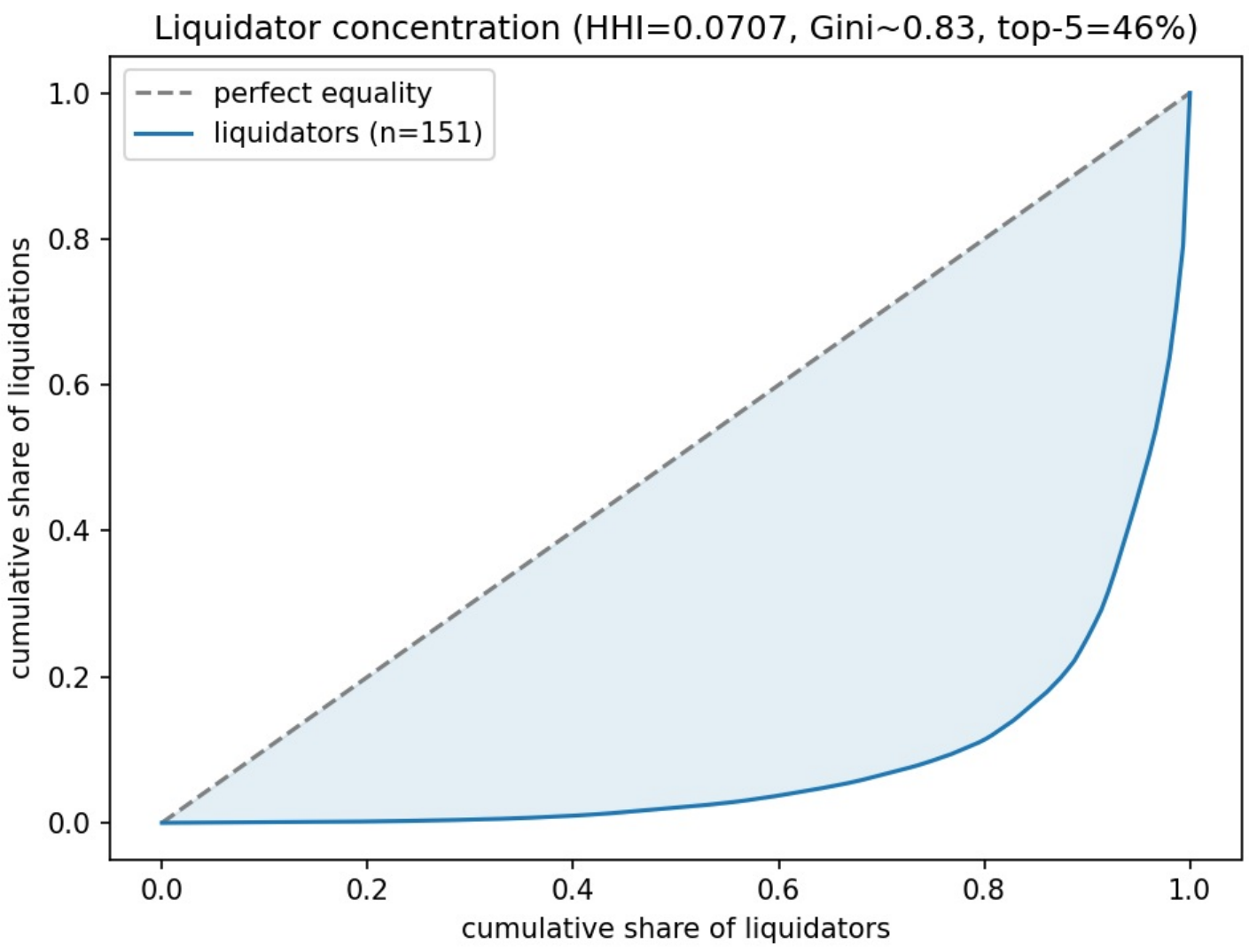

Figure 4c.1. Lorenz curve of liquidator concentration over the full record (HHI 0.071, Gini about 0.83, top-5 liquidators = 46 percent).

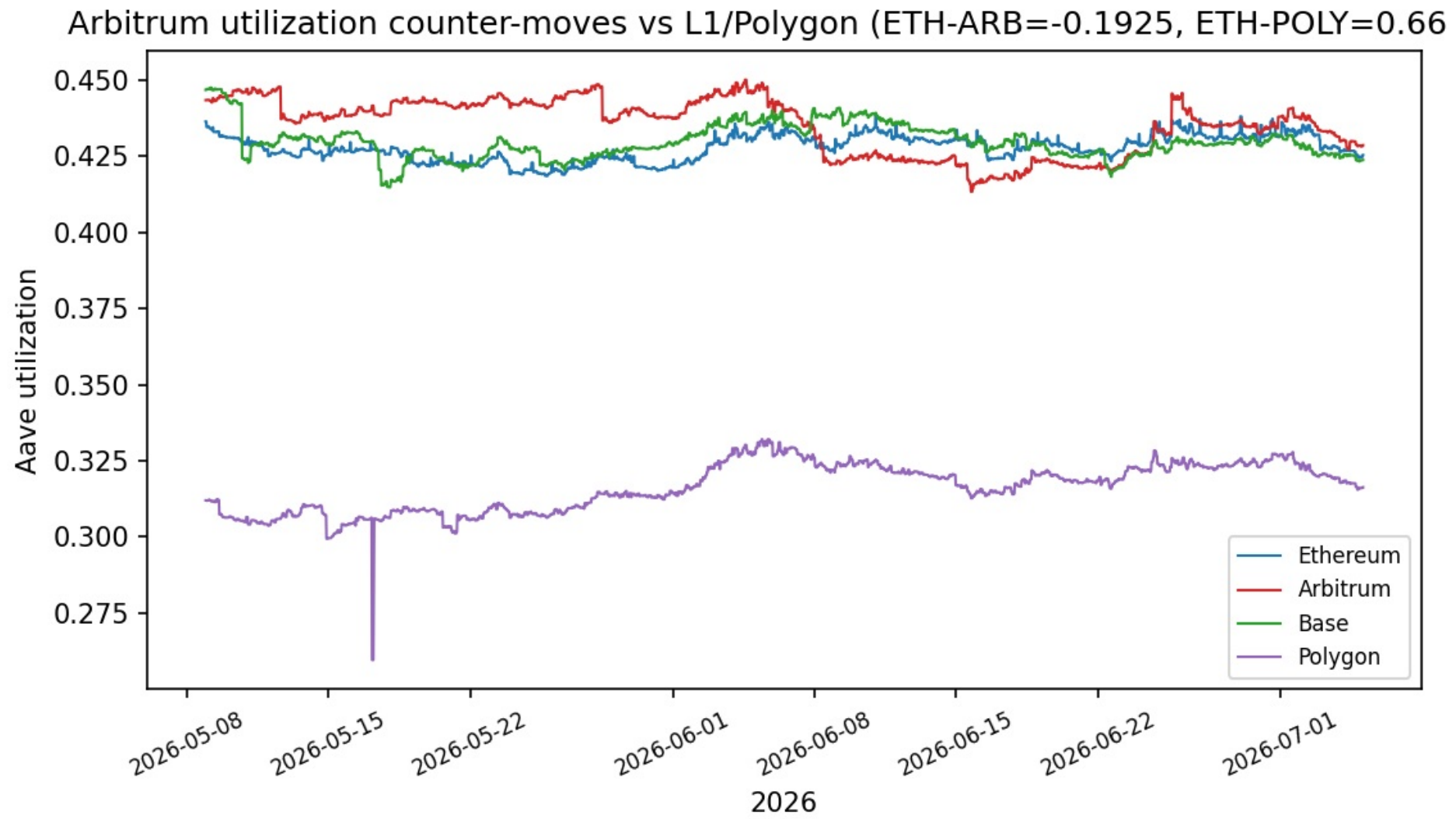

Figure 4c.2. Four-chain Aave utilization over the 2026 window. Arbitrum counter-moves against both Ethereum (correlation about −0.19) and Polygon (about −0.31), while Ethereum and Polygon co-move positively (about +0.67).

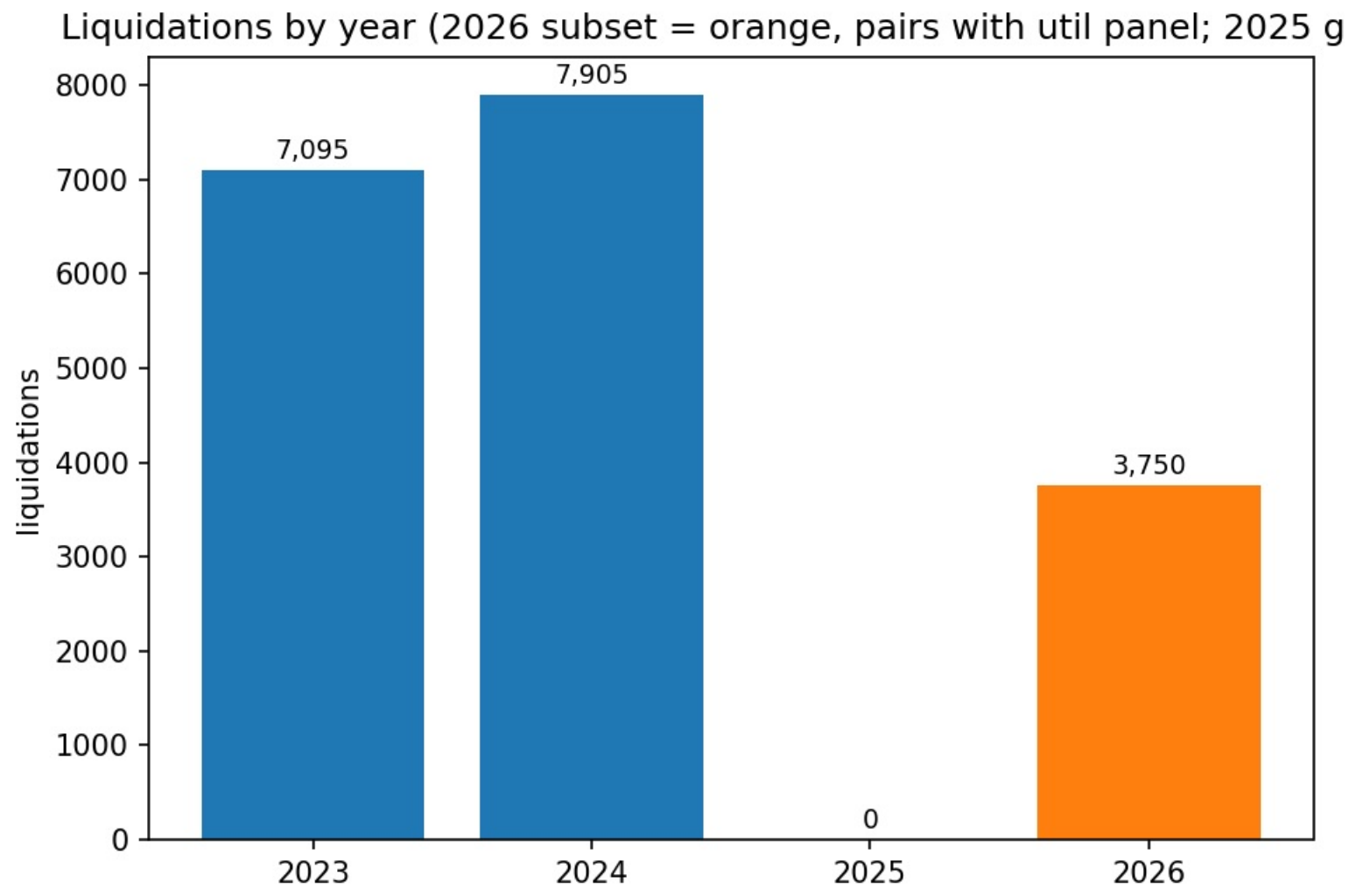

Figure 4c.3. Liquidations by year: 2023-2024 backfill, a 2025 gap, and the 2026 paired subset.

## 4d. Perpetual basis-funding coupling (RED-BASIS)

RED-BASIS is a 244,719-observation panel of spot-perpetual basis and eight-hour funding for BTC, ETH, and SOL on OKX, accompanied by a twenty-symbol funding cross-section and a BTC/ETH term-structure file. It is the corpus's window into the perpetual-derivatives layer, and it is the corpus's honest null. Funding and basis are mechanically linked, so one might expect the strength of their short-horizon coupling to carry information about impending funding-regime changes. We test this and find that it does not.

The one-hour rolling correlation between basis and funding hovers near zero for all three assets, with a mean of approximately 0.01 to 0.02 (Figure 4d.1). We then flag coupling breaks and ask whether a break predicts a funding-regime shift over the following four hours. Breaks raise the four-hour shift probability by only 2.9 to 4.5 percentage points over a base rate that is already near 91 percent, and the effect is statistically distinguishable from zero for only one of the three assets (Figure 4d.2). The near-certain base rate is itself the methodological lesson: a permissive regime-shift definition saturates the base rate, and the saturation grows with the horizon, from 43 percent at one hour to 97 percent at eight hours (Figure 4d.3). A predictor has little room to add information over a base rate that approaches one.

RED-BASIS matters to the corpus precisely because it does not exhibit structure. It is the layer where the concentration-and-one-sidedness motif does not appear, and it is reported as a null rather than suppressed. That contrast, structure in four layers and its honest absence in the fifth, is what keeps the Section 5 synthesis a finding rather than an artifact of how we chose to look.

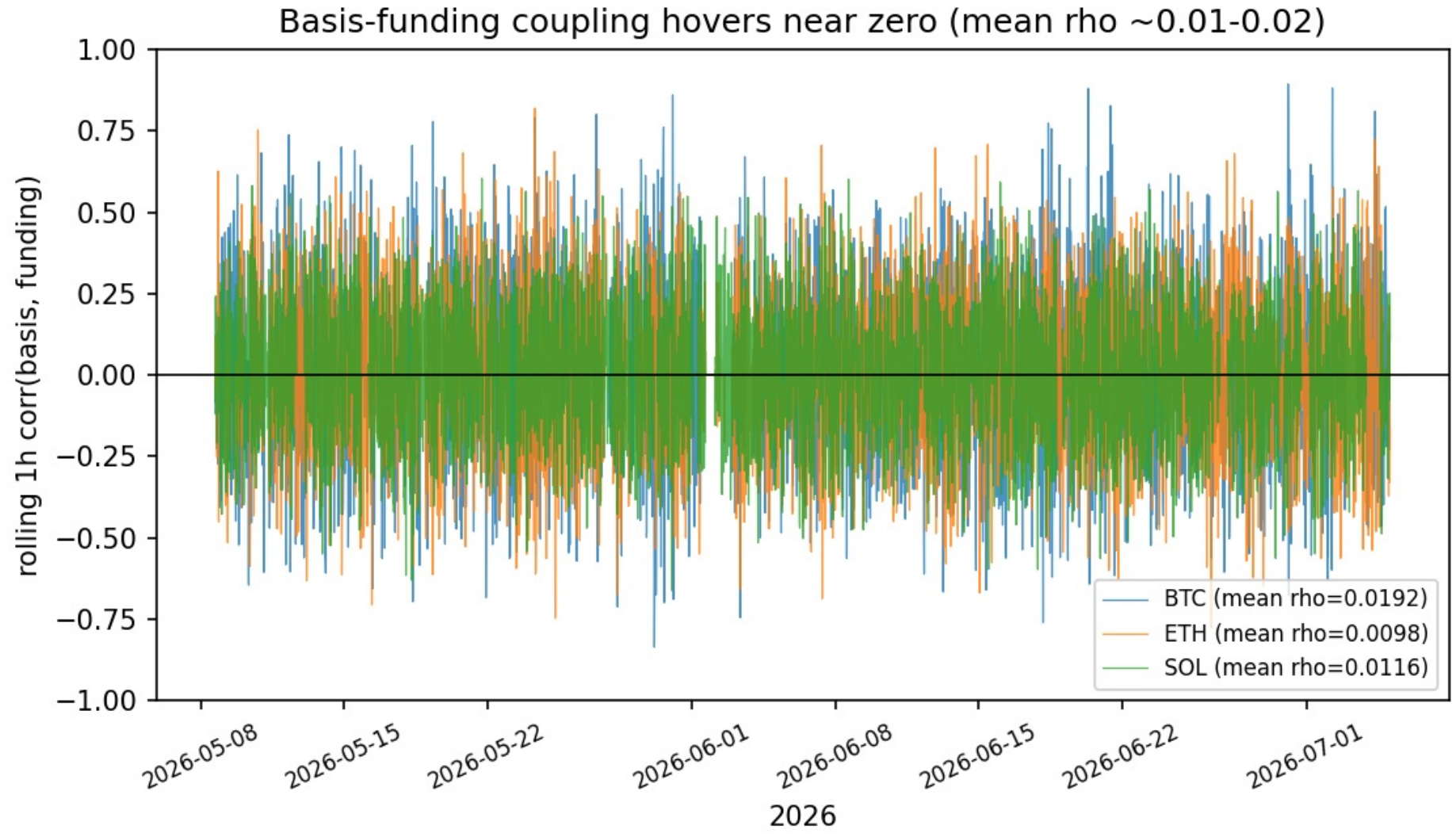

Figure 4d.1. Rolling one-hour correlation between basis and funding, per symbol; it hovers near zero.

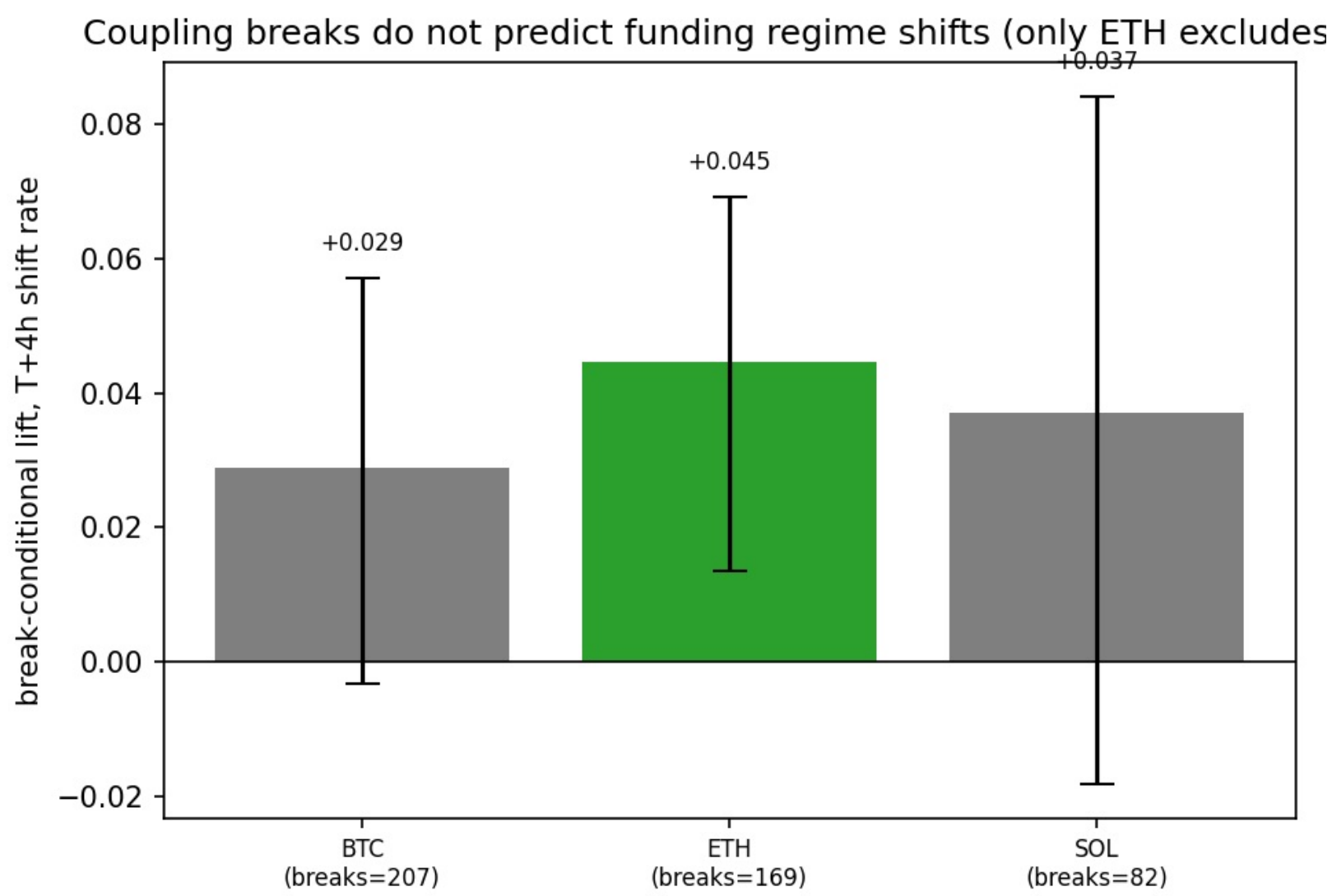

Figure 4d.2. Break-conditional lift in the four-hour regime-shift rate with 95 percent confidence intervals; only ETH excludes zero, and the effect is small.

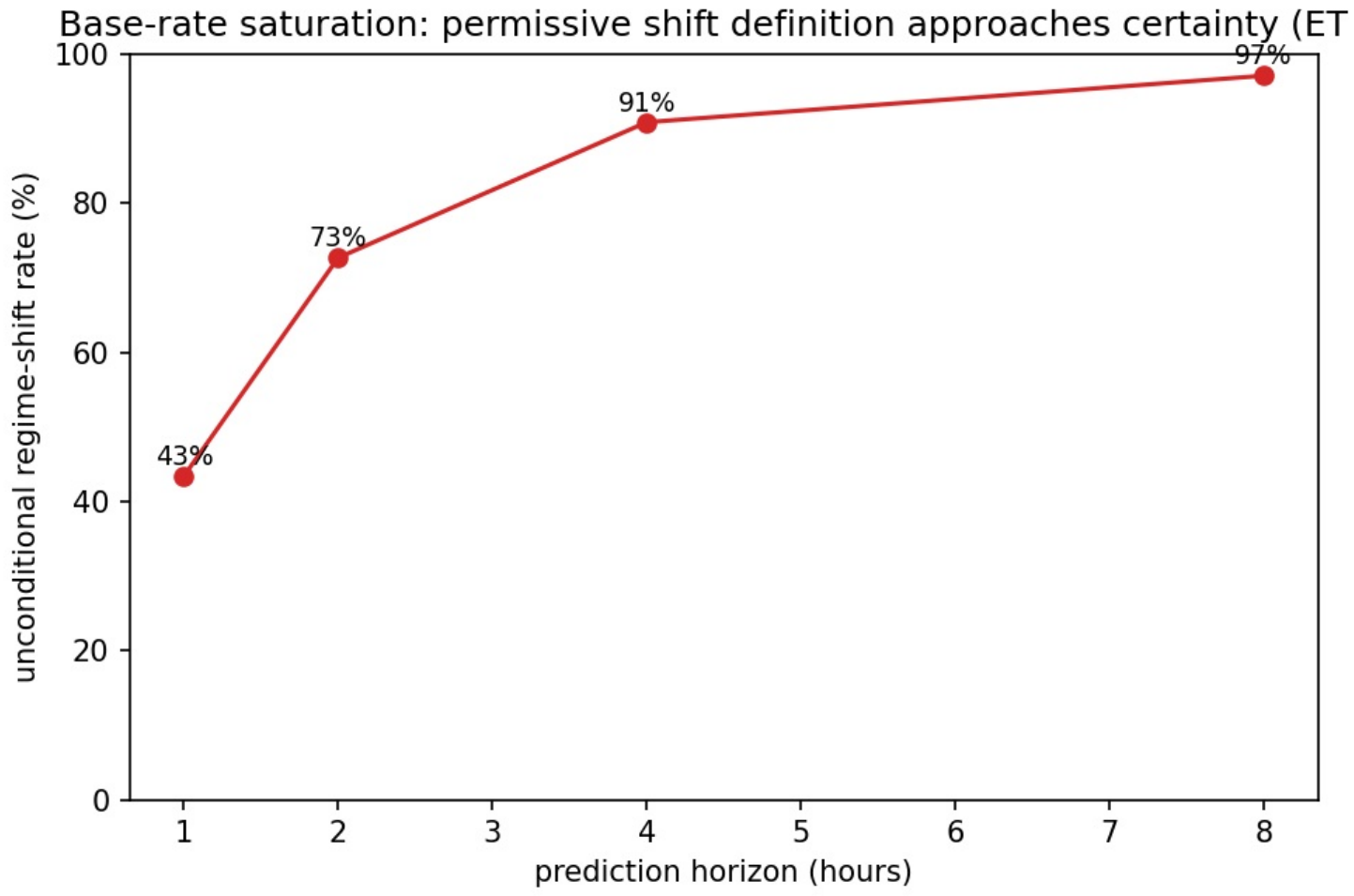

Figure 4d.3. Base-rate saturation: the unconditional regime-shift rate rises with horizon (43 percent at 1 hour to 97 percent at 8 hours).

## 4e. Wormhole cross-chain source-flow tape (RED-BRIDGE)

RED-BRIDGE is a 360,714-message tape of Wormhole cross-chain transfers, recording for each message a source chain, timestamps, transaction identifiers, and, where captured, a USD amount. It is the corpus's window into the cross-chain bridge layer. We are explicit at the outset about its completeness: the symbol field is unpopulated across all rows, the destination chain is present for only 29 percent of messages, and the status field is uniformly uninformative, so the analysis-ready fields are the source chain and the USD amount, and destination-side, per-asset, and status analyses are not supported by this release.

Within that boundary, two facts are clear. First, message origination is strongly concentrated and Solana-dominated: across 34 source chains, Solana originates 43.8 percent of messages, Ethereum 24.6 percent, and BSC 12.5 percent, for a source Herfindahl-Hirschman Index of 0.27 (Figure 4e.1). Second, the size distribution of the 96,240 USD-tagged messages is heavily retail-skewed: total tagged flow is 362.85 million dollars, but the median transfer is only 200 dollars, the 99th percentile is about 46,600 dollars, and the maximum is 8.85 million dollars (Figure 4e.3). Bridge usage in this window is dominated in count by small transfers, with aggregate value concentrated in a thin tail. Message flow by hour of day is also uneven (Figure 4e.2).

The Solana source-dominance is a third instance of the concentration motif, in yet another independent layer, and the retail-skewed size distribution is a further one-sidedness. We draw these together next.

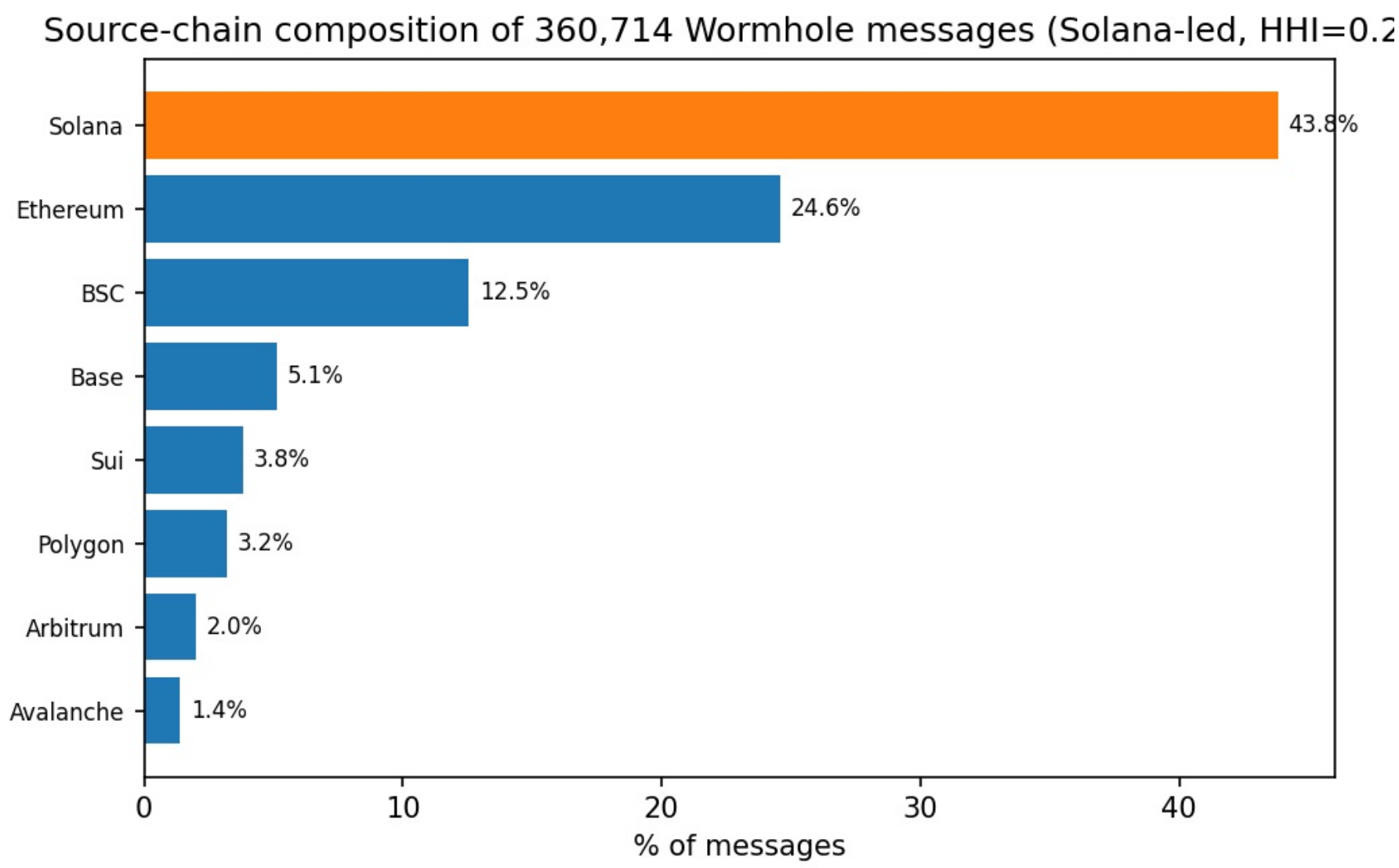

Figure 4e.1. Source-chain composition of the 360,714 Wormhole messages; Solana-led (source HHI 0.27).

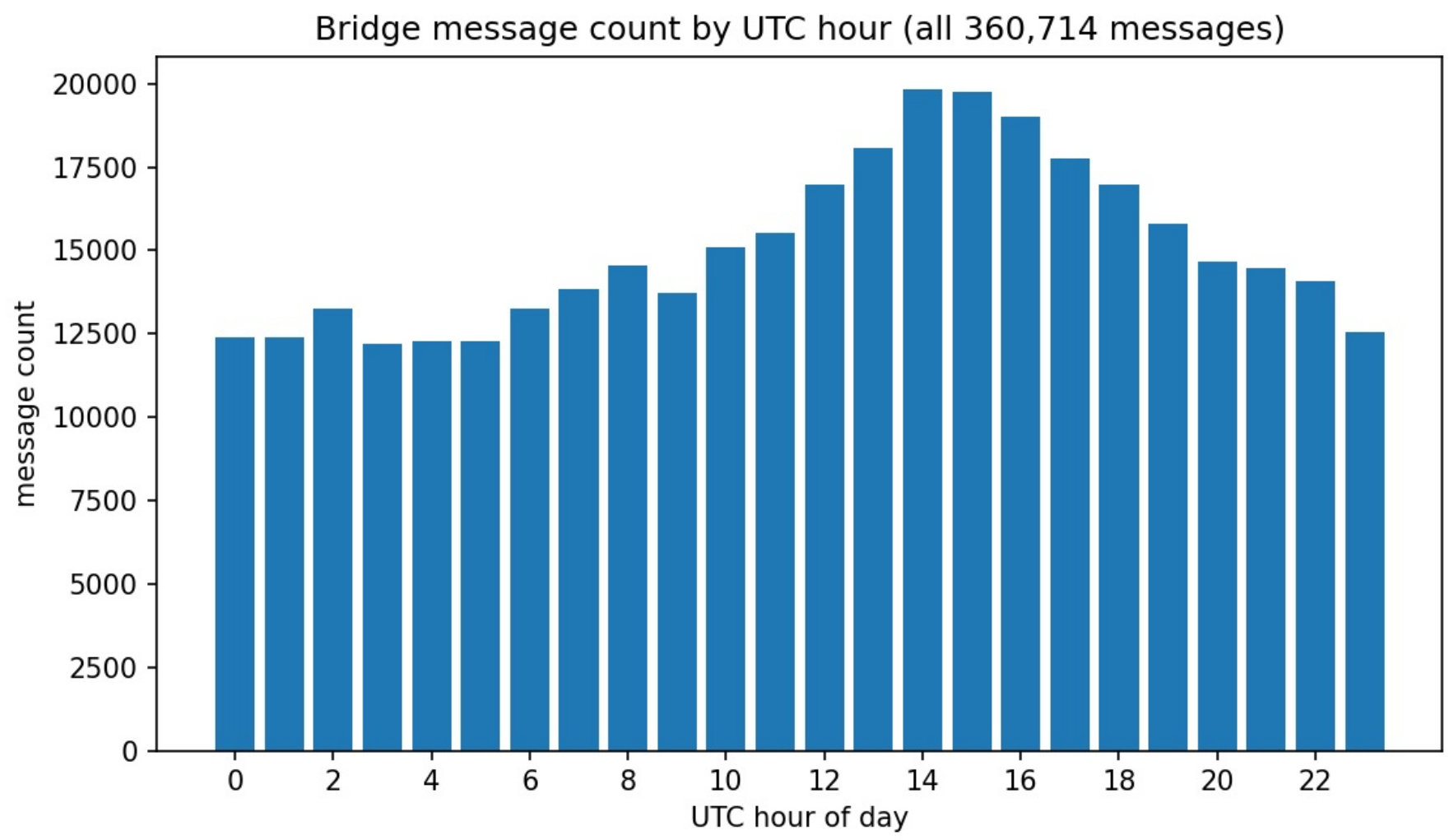

Figure 4e.2. Message count by UTC hour of day.

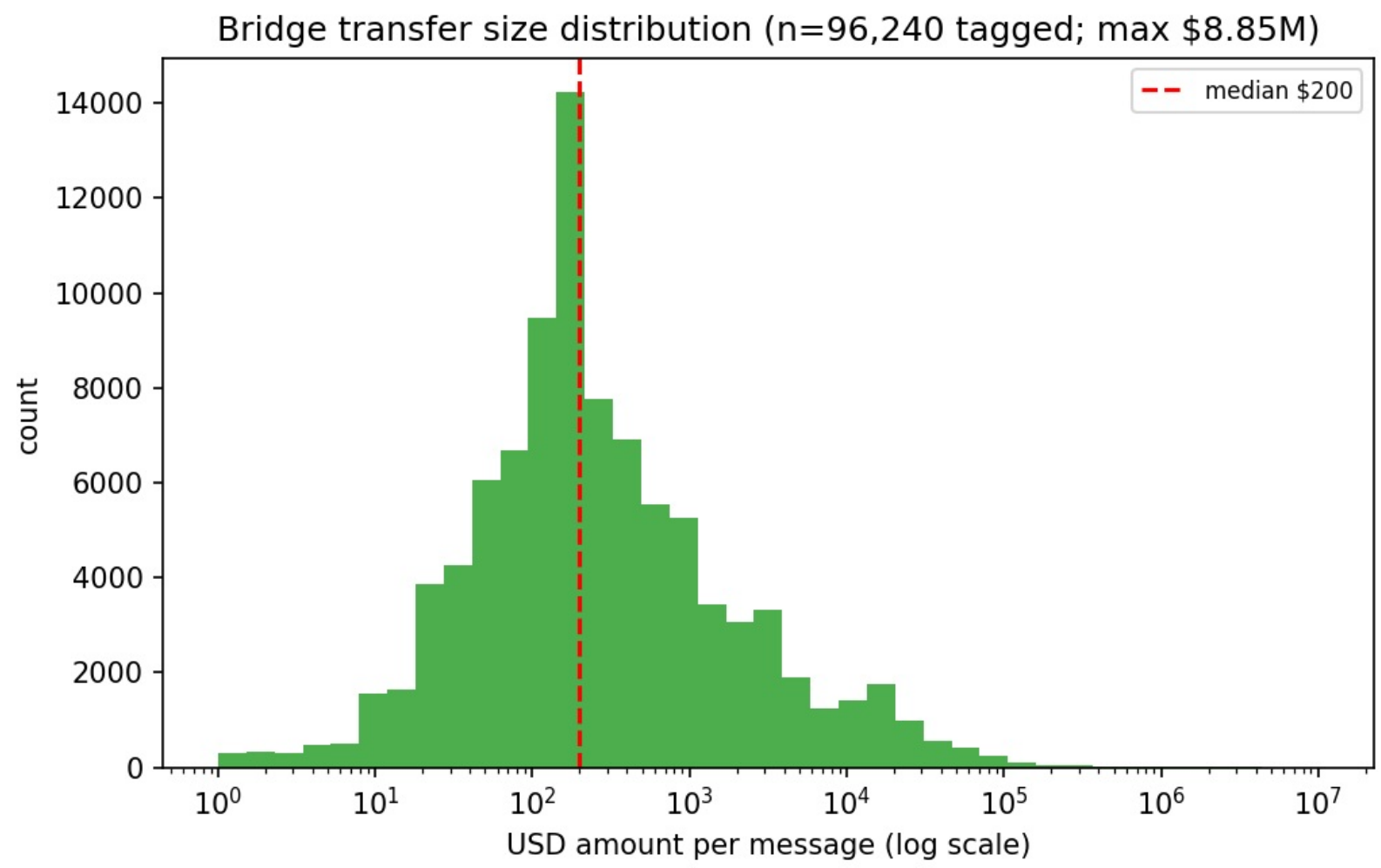

Figure 4e.3. USD transfer-size distribution (log scale) for the 96,240 USD-tagged messages; median 200 dollars, maximum 8.85 million dollars.

---

# 5. Cross-dataset discussion and synthesis

The five datasets were assembled to a common standard but measure five mechanically distinct phenomena, in five separate layers of the Solana and DeFi stack, using five separate data-collection pipelines. There is no reason a priori that a regularity found in one should appear in another. It is therefore notable that four of the five exhibit a common structural signature, and that the fifth does not. We describe that signature carefully, and we are deliberate about what we do and do not claim.

**A recurring signature: concentration and one-sidedness.** In four independent layers we observe the same two features. The first is concentration: a small number of actors, venues, or chains intermediate a disproportionate share of activity. In the spot layer, the winning-quote distribution across thirty-two Solana venues has a Herfindahl-Hirschman Index of 0.22, with the top five venues supplying 80 percent of best quotes and a single venue supplying 42 percent. In the lending layer, the liquidator distribution has a Gini coefficient of about 0.83 and a top-five share of 46 percent. In the bridge layer, message origination across thirty-four chains has a source Herfindahl-Hirschman Index of 0.27, with Solana alone originating 43.8 percent. The second feature is one-sidedness: directional offsets rather than symmetric dispersion around a center. In the spot layer, the decentralized price sits below the centralized benchmark in 91 percent of observations, a persistent offset rather than noise. In the oracle layer, drift grows with staleness in a single direction across lag deciles rather than fluctuating symmetrically. In the bridge layer, transfer sizes are heavily skewed toward a retail tail. The same two words, concentrated and one-sided, describe the spot tape, the liquidation record, the bridge tape, and the oracle panel, despite their having nothing mechanical in common.

**The honest exception.** The fifth dataset, RED-BASIS, does not exhibit this signature: the short-horizon coupling between basis and funding is near zero, and coupling breaks carry essentially no predictive content once the saturating base rate is accounted for. We report this null rather than suppress it, and its role in the corpus is important. If we had found structure in all five layers, a skeptical reader could reasonably suspect that the structure is an artifact of how we measure, a property of the observer rather than the market. The presence of a clean null in the fifth layer, produced by the same observer under the same methodology, is evidence that the signature in the other four is a property of the data and not of the lens.

**What we claim, and what we do not.** We do not claim a common cause. The concentration in the venue distribution, the concentration in the liquidator distribution, and the concentration in bridge origination are almost certainly driven by different mechanisms operating in different markets; asserting a single generative process would exceed what descriptive data can support. What we claim is narrower and, we think, more useful: that concentration and one-sidedness are empirically pervasive across independent Solana and DeFi microstructure layers in this window, that this pervasiveness is reproducible and quantified here, and that it is a hypothesis-generating observation rather than a tested mechanism. Candidate readings, offered as directions rather than findings, include the natural concentration of young markets with few sophisticated intermediaries, a Solana-ecosystem-specific intermediary structure, and the possibility that measurement against a centralized benchmark simply renders pre-existing asymmetries visible. Distinguishing among these requires an identification strategy the present corpus is designed to enable but does not itself provide.

**Relationship to the Ethereum canon.** The canon established that decentralized venues are less price-efficient than centralized ones and that gas frictions are a principal driver (Barbon and Ranaldo 2024; Lehar and Parlour 2025; Capponi, Jia, and Yu forthcoming). Our corpus does not contest that result; it complements it by characterizing, on Solana specifically and across adjacent layers, the shape of the inefficiency: not diffuse noise but concentrated intermediation and one-sided offsets. Read alongside the canon, the corpus suggests that the interesting question for Solana microstructure is less whether it is efficient than how its inefficiency is structured.

**Reproducibility infrastructure as a contribution.** Finally, we regard the reproducibility discipline itself, one observer, one window, one methodology, fixed seeds, deposited scripts and results, integrity manifests, and a permissive licence, as a first-class contribution independent of any single finding. It is what makes the four-layer signature comparable in the first place, and it is what allows a later researcher to test the hypotheses above against the same data without re-deriving it.

# 6. Limitations

The corpus is descriptive and single-window, and each dataset carries specific limits that bound its use. RED-ORACLE admits no valid instrument, because publish lag is definitionally the difference between sample and publish times; its staleness-drift relationship is an association only. RED-ARB is an observation tape without signed trades or per-block liquidity depth, so it supports neither execution-quality nor information-based (PIN/VPIN) claims, and its spreads are not realizable arbitrage. RED-LIQ spans multiple years with a 2025 gap, so only the approximately 3,750 liquidations dated 2026 pair with the utilization panel; cross-panel statements are restricted to that subset. RED-BASIS reports a null that is conditional on the regime-shift definition, and we show explicitly that permissive definitions saturate the base rate, so the null should be read as “no predictive content under reasonable definitions” rather than a universal claim. RED-BRIDGE has unpopulated symbol, destination, and status fields, restricting it to source-side flow and USD size. All five were collected from public and free-tier feeds, which favors reproducibility and independence but forgoes the depth of paid or exchange-internal data; all cover a single continuous 2026 window of approximately 57 days, so seasonality and regime dependence are untested; and small-count features (for example the 21 oracle episodes above 60 seconds, and the 2,694-observation utilization panel) are flagged as such and should not bear heavy inference.

## 7. Conclusion

We have released and documented the RED-2400 family v2, a corpus of five public Solana and DeFi microstructure datasets built to a single reproducibility standard, and we have reported their descriptive contents honestly: three substantive findings, one null, and one field-completeness characterization. Drawing the five together, we observed a recurring signature of concentration and one-sidedness across four independent layers, sharpened rather than undermined by its clean absence in the fifth, and we framed that signature as a reproducible, hypothesis-generating observation rather than a tested mechanism. The natural next step is causal identification the present corpus enables but does not supply: for example, a study of whether the persistent DEX-below-CEX offset compresses with venue entry or over time, which would move the spot-layer observation from description toward mechanism. Longer-horizon and monthly-versioned re-releases would test the window dependence noted above, and cross-family analysis, joining these datasets to external corpora such as the Aave infrastructure of Fan and Sun (2025), would extend the benchmark's reach. Our aim here is narrower and prior to all of that: to lower the data barrier, and to put a coherent, reproducible, honestly documented benchmark family in the hands of the researchers who will ask those questions.

---

## 8. Data availability

All five datasets in the RED-2400 family v2 are deposited on Zenodo under the CC-BY-4.0 licence and are independently citable via their DOIs:

- RED-ORACLE: `10.5281/zenodo.21211110`
- RED-LIQ: `10.5281/zenodo.21211303`
- RED-BASIS: `10.5281/zenodo.21211415`
- RED-ARB: `10.5281/zenodo.21211187`
- RED-BRIDGE: `10.5281/zenodo.21211498`

Each deposit includes the primary data files, a fixed-seed reproducibility script, a SHA-256 integrity manifest, complete field-level schema documentation, and a companion `README`. Full per-dataset schemas and detailed reproducibility tables are provided in each Zenodo deposit (see Section 3).

---

*Reference-verification note: All references above were independently verified against the primary source (journal DOI, arXiv record, SSRN abstract, or working-paper number) before inclusion.*

## LLM acknowledgment

Preliminary manuscript draft prepared with assistance from an Anthropic Claude LLM session. All datasets, numerical results, methodology, and scientific framing are the author’s own.